\documentclass[
reprint,
superscriptaddress,
amsmath,
amssymb,
aps,
prb,
twocolumn,
showpacs]{revtex4-1} 

\usepackage{graphicx}
\graphicspath{{../}}
\usepackage[utf8]{inputenc}
\usepackage[english]{babel}
\usepackage{color}
\usepackage{soul}
\usepackage[section]{placeins}
\usepackage{tabularx}
\usepackage{array}
\usepackage[separate-uncertainty=true]{siunitx}
\usepackage{xcolor}
\usepackage{bm}
\usepackage{hyperref}
\hypersetup{
   colorlinks=true,     
   linkcolor=blue,      
   citecolor=blue,      
   filecolor=blue,   
   urlcolor=blue       
}

\newcommand{\ms}{\ensuremath{\mathrm{M_{S}}}}
\newcommand{\keff}{\ensuremath{\mathrm{K}_{\mathrm{eff}}} }
\newcommand{\ku}{\ensuremath{\mathrm{K_{u}}} }

\newcommand{\hc}{\ensuremath{\mathrm{H_{C}}}}
\newcommand{\xTb}{\ensuremath{t_\mathrm{Tb}}}
\newcommand{\hsat}{\ensuremath{\mathrm{H_{Sat}}}}

\DeclareSIUnit\angstrom{\text {Å}}
\DeclareSIUnit\bar{bar}
\begin{document}

\title{Perpendicularly magnetized Tb/Co multilayers featuring tilted uniaxial anisotropy: Experiments and modeling}

\author{J. C. Rodriguez E.}
\affiliation{Departamento Magnetismo y Materiales Magnéticos, Gerencia de Física, Centro Atómico Bariloche, Av. E. Bustillo 9500 (R8402AGP), San Carlos de Bariloche, R\'{\i}o Negro, Argentina.}
\affiliation{Instituto de Nanociencia y Nanotecnolog\'{\i}a, (CNEA--CONICET), Nodo Bariloche, Av. E. Bustillo 9500 (R8402AGP), San Carlos de Bariloche, R\'{\i}o Negro, Argentina.}
\affiliation{Instituto Balseiro, Universidad Nacional de Cuyo--CNEA, Av. E. Bustillo 9500 (R8402AGP) San Carlos de Bariloche, R\'{\i}o Negro, Argentina.}
\author{L. Avilés-Félix}
\affiliation{Departamento Magnetismo y Materiales Magnéticos, Gerencia de Física, Centro Atómico Bariloche, Av. E. Bustillo 9500 (R8402AGP), San Carlos de Bariloche, R\'{\i}o Negro, Argentina.}
\affiliation{Instituto de Nanociencia y Nanotecnolog\'{\i}a, (CNEA--CONICET), Nodo Bariloche, Av. E. Bustillo 9500 (R8402AGP), San Carlos de Bariloche, R\'{\i}o Negro, Argentina.}
\affiliation{Instituto Balseiro, Universidad Nacional de Cuyo--CNEA, Av. E. Bustillo 9500 (R8402AGP) San Carlos de Bariloche, R\'{\i}o Negro, Argentina.}
\author{M. H. Aguirre}
\affiliation{Instituto de Nanociencia y Materiales de Aragón (INMA-CSIC), Campus Río Ebro, Universidad de Zaragoza, Spain.}
\affiliation{Dpto. de Física de la Materia Condensada, Campus Río Ebro, Universidad de Zaragoza, Spain.}
\affiliation{Laboratorio de Microscopías Avanzadas Edificio I+D, Campus Río Ebro, Universidad de Zaragoza, Spain.}
\author{L. M. Rodriguez}
\affiliation{Departamento de Interacción de la Radiación con la Materia, Gerencia de Física, Centro Atómico Bariloche, Av. E. Bustillo 9500 (R8402AGP), San Carlos de Bariloche, R\'{\i}o Negro, Argentina.}
\affiliation{Instituto de Nanociencia y Nanotecnolog\'{\i}a, (CNEA--CONICET), Nodo Bariloche, Av. E. Bustillo 9500 (R8402AGP), San Carlos de Bariloche, R\'{\i}o Negro, Argentina.}
\author{D. Salomoni}
\affiliation{Spintec, Université Grenoble Alpes, CNRS, CEA, Grenoble INP, IRIG-SPINTEC, 38000 Grenoble, France.}
\author{S. Auffret}
\affiliation{Spintec, Université Grenoble Alpes, CNRS, CEA, Grenoble INP, IRIG-SPINTEC, 38000 Grenoble, France.}
\author{R. C. Sousa}
\affiliation{Spintec, Université Grenoble Alpes, CNRS, CEA, Grenoble INP, IRIG-SPINTEC, 38000 Grenoble, France.}
\author{I. L. Prejbeanu}
\affiliation{Spintec, Université Grenoble Alpes, CNRS, CEA, Grenoble INP, IRIG-SPINTEC, 38000 Grenoble, France.}
\author{A. E. Bruchhausen}
\affiliation{Departamento de Fotonica y Optoelectrónica, Gerencia de Física, Centro Atómico Bariloche, Av. E. Bustillo 9500 (R8402AGP), San Carlos de Bariloche, R\'{\i}o Negro, Argentina.}
\affiliation{Instituto de Nanociencia y Nanotecnolog\'{\i}a, (CNEA--CONICET), Nodo Bariloche, Av. E. Bustillo 9500 (R8402AGP), San Carlos de Bariloche, R\'{\i}o Negro, Argentina.}
\affiliation{Instituto Balseiro, Universidad Nacional de Cuyo--CNEA, Av. E. Bustillo 9500 (R8402AGP) San Carlos de Bariloche, R\'{\i}o Negro, Argentina.}
\author{E. De Biasi}
\affiliation{Departamento Magnetismo y Materiales Magnéticos, Gerencia de Física, Centro Atómico Bariloche, Av. E. Bustillo 9500 (R8402AGP), San Carlos de Bariloche, R\'{\i}o Negro, Argentina.}
\affiliation{Instituto de Nanociencia y Nanotecnolog\'{\i}a, (CNEA--CONICET), Nodo Bariloche, Av. E. Bustillo 9500 (R8402AGP), San Carlos de Bariloche, R\'{\i}o Negro, Argentina.}
\affiliation{Instituto Balseiro, Universidad Nacional de Cuyo--CNEA, Av. E. Bustillo 9500 (R8402AGP) San Carlos de Bariloche, R\'{\i}o Negro, Argentina.}
\author{J. Curiale}
\email{jcuriale@cnea.gob.ar, he/him/his}
\affiliation{Departamento Magnetismo y Materiales Magnéticos, Gerencia de Física, Centro Atómico Bariloche, Av. E. Bustillo 9500 (R8402AGP), San Carlos de Bariloche, R\'{\i}o Negro, Argentina.}
\affiliation{Instituto de Nanociencia y Nanotecnolog\'{\i}a, (CNEA--CONICET), Nodo Bariloche, Av. E. Bustillo 9500 (R8402AGP), San Carlos de Bariloche, R\'{\i}o Negro, Argentina.}
\affiliation{Instituto Balseiro, Universidad Nacional de Cuyo--CNEA, Av. E. Bustillo 9500 (R8402AGP) San Carlos de Bariloche, R\'{\i}o Negro, Argentina.}
\date{\today}

\begin{abstract}
Rare earth/transition metal (RE/TM) multilayers with perpendicular magnetic anisotropy are key ingredients for the development of spintronic applications. Their compensation temperature depends on the ratio of the thicknesses of rare earth and transition metal, allowing their magnetic properties to be tuned with temperature while maintaining their anisotropy even in nanometer-scale devices.
In this work, we performed a thorough structural characterization and systematically investigate the magnetic properties of a whole family of ferrimagnetic [Tb/Co]$_{\times 5}$ multilayers varying the Tb thickness in the range of \qty{0.4}{\nm} - \qty{1.25}{\nm}.
A linear dependence of the compensation temperature on the Tb layer thickness was observed.
Moreover, a uniaxial anisotropy constant of $\qty{330+-30}{\unit{\kilo\J}/\unit{\m}^3}$, which is close to the values reported by other authors, was estimated. 
Additionally, we proposed a model to gain a better understanding of the angular dependence of the magnetization loops and the linear dependence of the compensation temperature. We present strong evidence demonstrating that the perpendicular anisotropy must be tilted away from the perpendicular axis in order to explain the observed features, particularly the hysteresis in the in-plane loops.
Our work advances the understanding of DC magnetic properties in thin RE/TM ferrimagnetic films, which has the potential to impact different fields where these materials are involved. 
\end{abstract}

\maketitle

\section{Introduction}
Rare earth (RE)-transition metal (TM) alloys and multilayers are undergoing significant development due to their promising applications; particularly their strong perpendicular magnetic anisotropy (PMA) \cite{kim2022, Tsunashima2001} which is difficult to achieve in single TM ferromagnets \cite{DenBroeder1991}. Additionally, ferrimagnets have the possibility  to tune its anisotropy constant by varying the relative concentration between the RE and the TM \cite{Hansen1989}. These features are extremely valuable for the development of perpendicular magnetic tunnel junctions (p-MTJ), which require storage and reference layers with strong PMA at low nanometer-scale lateral dimensions.

In ferrimagnetic multilayers, the RE and the TM sublattices are antiferromagnetically coupled at the interface; however the material exhibits a finite total magnetization. As known, the magnetization of ferrimagnets vanishes at a particular temperature, which is referred to as magnetic compensation temperature $\mathrm{T_{M}}$, and depends on the relative RE and TM composition.
The existence of the $\mathrm{T_{M}}$ enables a thorough tuning of the magnetic properties
for the development of different spintronic devices with enhanced functionalities; e.g. all-optical switching (AOS) of magnetization \cite{LAvilesSR2020, Iihama2018, Mondal2023}, spin-orbit torque-induced switching \cite{Je2018}, high-frequency oscillators, \cite{HaidarNC2019} magneto-optical recording \cite{Tsunashima2001}, etc; being another advantage over traditional ferromagnets.

In this framework, TbCo multilayers and alloys have gained attention in the recent years \cite{Ciuciulkaite2020, Frackowiak2022}. For instance, a clear linear dependence of the $\mathrm{T_M}$ with RE  composition was reported \cite{Alebrand2012, LAvilesSR2021, Nava2025}, but no clear connection to the magnetic properties has been established.
Further works report on samples with large PMA that exhibit in-plane hysteresis \cite{ErtlJMMM1992,Gottwald2012, Mishra2023}, but this behavior is rather unexpected for a uniaxial magnetic system with a dominant PMA. In addition, although the magnetization switching in Tb/Co has been widely studied during the last years with special focus on AOS, there are several issues dealing with the magnetic anisotropies that remains unclear. Some authors \cite{Mishra2023, Peng2023} attribute the observed behavior to an in-plane anisotropy, but a systematic and comprehensive study of the magnetic properties as a function of thickness and temperature in Tb/Co multilayers has not been reported.
It is also important to note that such anisotropies are effective anisotropies, and therefore do not necessarily coincide with the local microscopic anisotropies that give rise to the sperimagnetic behavior observed in RE–TM alloys \cite{Hu_SperimagneticAmAll_PRB2025}, nor with the conical magnetic configurations typically formed by the RE moments in those systems \cite{DrovosekovTwistedM_Multilayers_JMMM2019}.

In the present work, we study magnetic properties of a complete set of ferrimagnetic Tb/Co multilayers with varying Tb thickness on the Co-rich side. A linear relationship between remanent magnetization and temperature was consistently observed across all samples, allowing us to propose a straightforward model to predict the compensation temperature $\mathrm{T_{M}}$. We also develop a toy model to reproduce the hysteresis cycle considering the magnetization reversal by domain nucleation. The paper is organized as follows: In Sec. II we present results related to the structural and chemical characterization of the Tb/Co multilayers. In Sec. III we discuss the magnetic properties of the samples, highlighting the influence of the Tb thickness on the compensation temperature and other relevant parameters, such as saturation magnetization, coercive field, anisotropies, etc. In this section, the developed magnetic model for the system is also presented, indicating successes and constraints. Finally, Sec. IV is devoted to summarize our work.

\section{Structural Characterization}
 
\begin{figure}[htbp]
\centering
\includegraphics[width=\columnwidth]{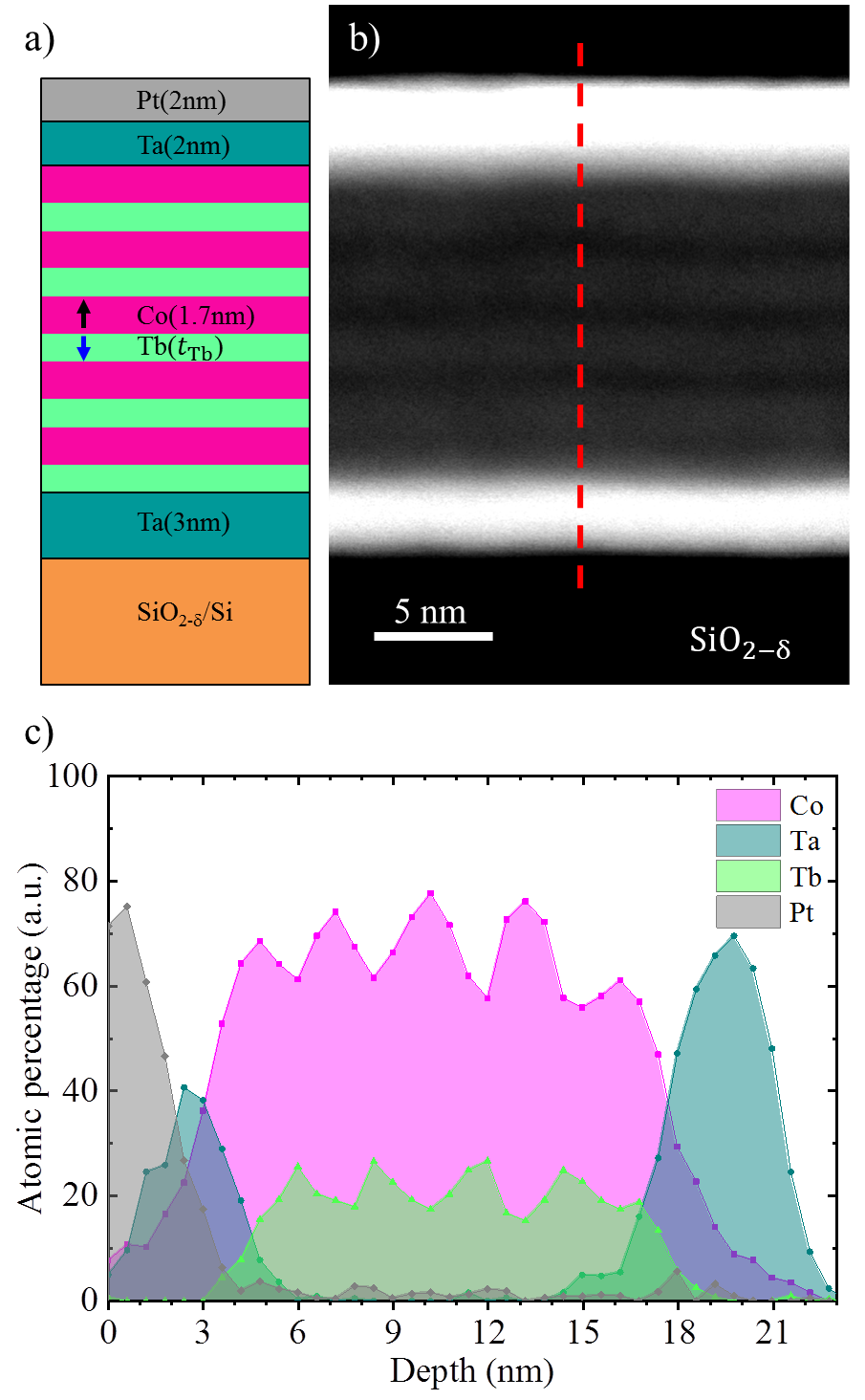}
\caption{\label{fig:SampleStructure}
Sample structure and composition profile. (a) Schematic of ferrimagnetic multilayers. (b) HAADF-STEM cross-sectional image for an TbCo(\qty{1.08}{\nm}) multilayer (c) EDX map plotted with depth of \qty{20}{\nm} thick specimen indicated by the red dashed line in panel (b).} 
\end{figure}

The studied samples are multilayers consisting of Ta(\qty{3}{\nm}) / [Tb(\xTb) / Co(\qty{1.7}{\nm}))]$_{\times 5}$ / Ta(\qty{2}{\nm}) /Pt(\qty{2}{\nm}) multilayers as schematized in Fig. \ref{fig:SampleStructure}a). The thickness of each layer is indicated in parenthesis. The square brackets contain the terbium-cobalt bilayer structure that is repeated \num{5} times. Through the manuscript, the label TbCo(\xTb) will be used as notation for samples with different nominal Tb thickness (\xTb), given in \unit{\nm}. 

The multilayers were grown on a 4-inch thermally-oxidized silicon wafer (SiO$_2\approx$ \qty{100}{\nm}) by DC magnetron sputtering using an Ar pressure of \qty{2}{\milli\bar} and a base pressure of \qty{e-8}{\milli\bar} \cite{LAvilesSR2020}.
The sample was grown with a constant layer thickness for all metals throughout the wafer, with the exception of the Tb layers.
For this last metal, the growth was performed off-axis, establishing a thickness gradient along a direction perpendicular to the wafer flat. In this direction, the Tb thickness was varied in the range $\qty{0.4}{\nm} <$ \xTb~$< \qty{1.25}{\nm}$. The 4-inch silicon wafer was cut into pieces of $\qty{1}{\cm}\times \qty{1}{\cm}$ using an automatic dicing saw. The label of the samples ``\xTb'' represents the nominal Tb thickness at the center of each sample. Note that the thickness of each Tb layer changes considerably less than one lattice parameter within each of the \qty{1}{\cm^2} pieces.

The structure of multilayers was studied by high-angle annular dark-field scanning transmission electron microscopy (HAADF-STEM) and  energy dispersive X-ray spectroscopy (EDX) with a resolution of approximately \qty{3}{\angstrom}. 
Lamella-shaped samples were thinned and transferred using the Focused Ion Beam (FIB) technique.
In Fig. \ref{fig:SampleStructure}b), a characteristic HAADF-STEM micrograph of the lamella is shown, corresponding to sample TbCo(\qty{1.08}{\nm}). As the gray level is proportional to the atomic number $\mathrm{Z}^{1.8}$, the Tb and Co layers are clearly distinguishable. The red dashed line indicates the position of the EDX profile shown in Fig. \ref{fig:SampleStructure}c). 
In this case, as it was expected, five distinct peaks of Tb and Co are observed. Although interdiffusion cannot be neglected, the fact that in the position of maximum Co contents the percentage of Tb is not zero, and vice versa, usually indicates the presence of interfacial roughness.

\begin{figure}[htbp]
\centering
\includegraphics[width=\columnwidth]{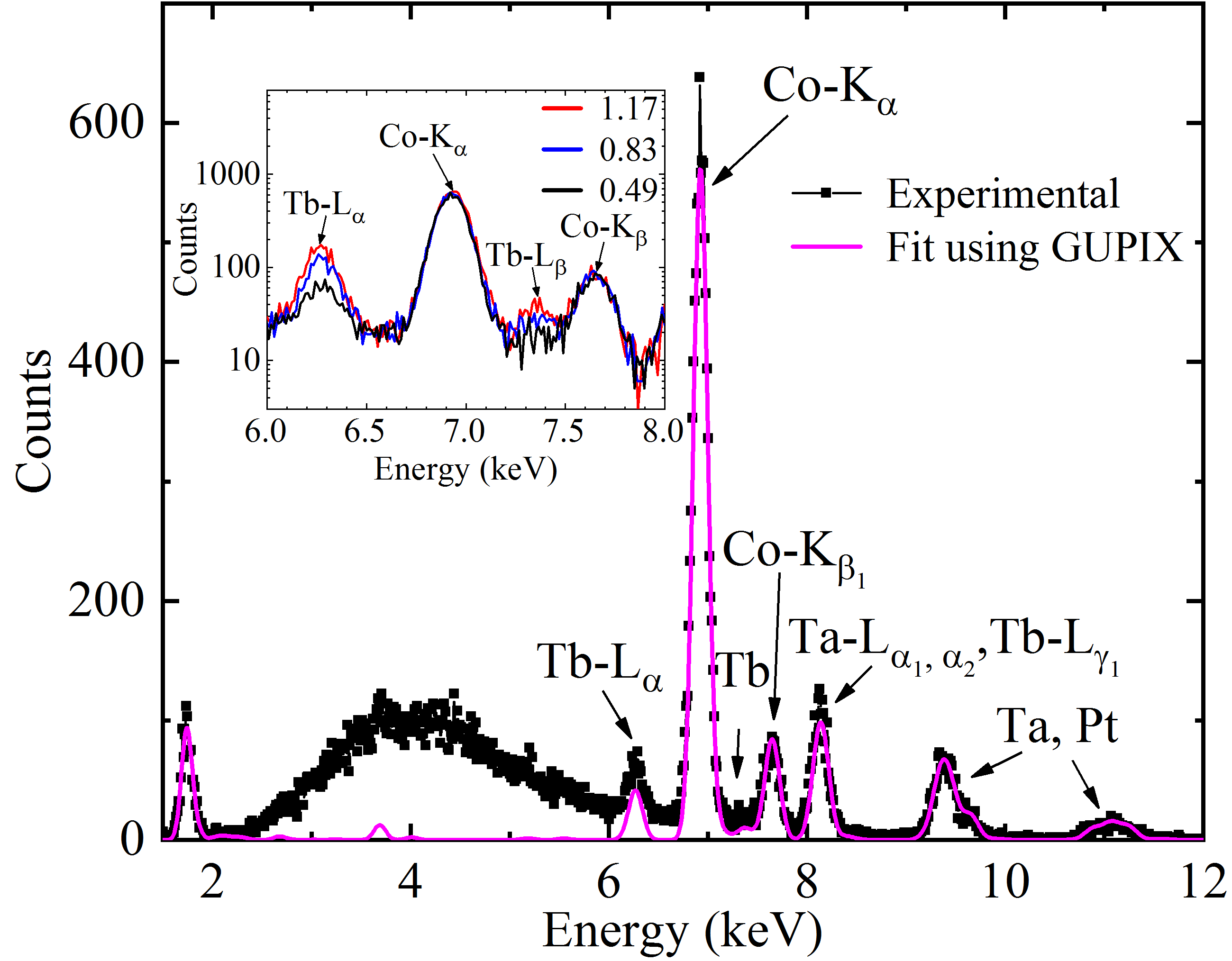}
\caption{\label{fig:SamplePIXEandRBS}
The experimental PIXE spectrum for \qty{3}{\mega \electronvolt}  $H^{+}$ of the TbCo(\qty{0.49}{\nm}) multilayer (black squares and solid line)  using Mylar as outgoing X-rays filter. The magenta solid line represents the best fit provided by the GUPIX code. The inset shows the experimental PIXE spectra of TbCo(\xTb) multilayer highlighting the Tb and Co peaks between \num{6} and \qty{8}{\kilo \electronvolt} for \xTb= \num{0.49}, \num{0.83} and \num{1.17}, with black, blue and red lines, respectively.
}
\end{figure}

TbCo(\xTb) multilayers were also characterized by Particle-Induced X-ray Emission (PIXE). Characteristic PIXE spectra are shown in Fig. \ref{fig:SamplePIXEandRBS}, where peaks corresponding to Tb, Co, Ta, Pt and Si elements were identified. A Mylar filter was used during measurements to block outgoing X-rays below \qty{4} {\kilo \electronvolt} in order to reduce the high yield coming from the Si wafer at \qty{1.74} {\kilo \electronvolt}, and therefore to improve the detection of Tb and Co by better defining their contributions. In addition, the inset in Fig. \ref{fig:SamplePIXEandRBS} displays, in a semi-logarithm scale, a zoom of the PIXE spectra for \xTb=\num{0.49}, \num{0.83} and \num{1.17}. The amplitude of Tb peaks increase as a function of \xTb, while the intensity of Co peaks remains practically constant. A quantitative analysis of these spectra using the GUPIX code \cite{Campbell2010} gives the Tb-Co mass fraction to be of \num{19}\% (\num{81}\%), \num{36}\% (\num{64}\%) and \num{43}\% (\num{57}\%), respectively, in good agreement with the expected results for nominal thicknesses.

\section{Magnetic Characterization}
\subsection{Influence of the Tb thickness on the magnetic properties}
\begin{figure}[htbp]
\centering
\includegraphics[width=\columnwidth]{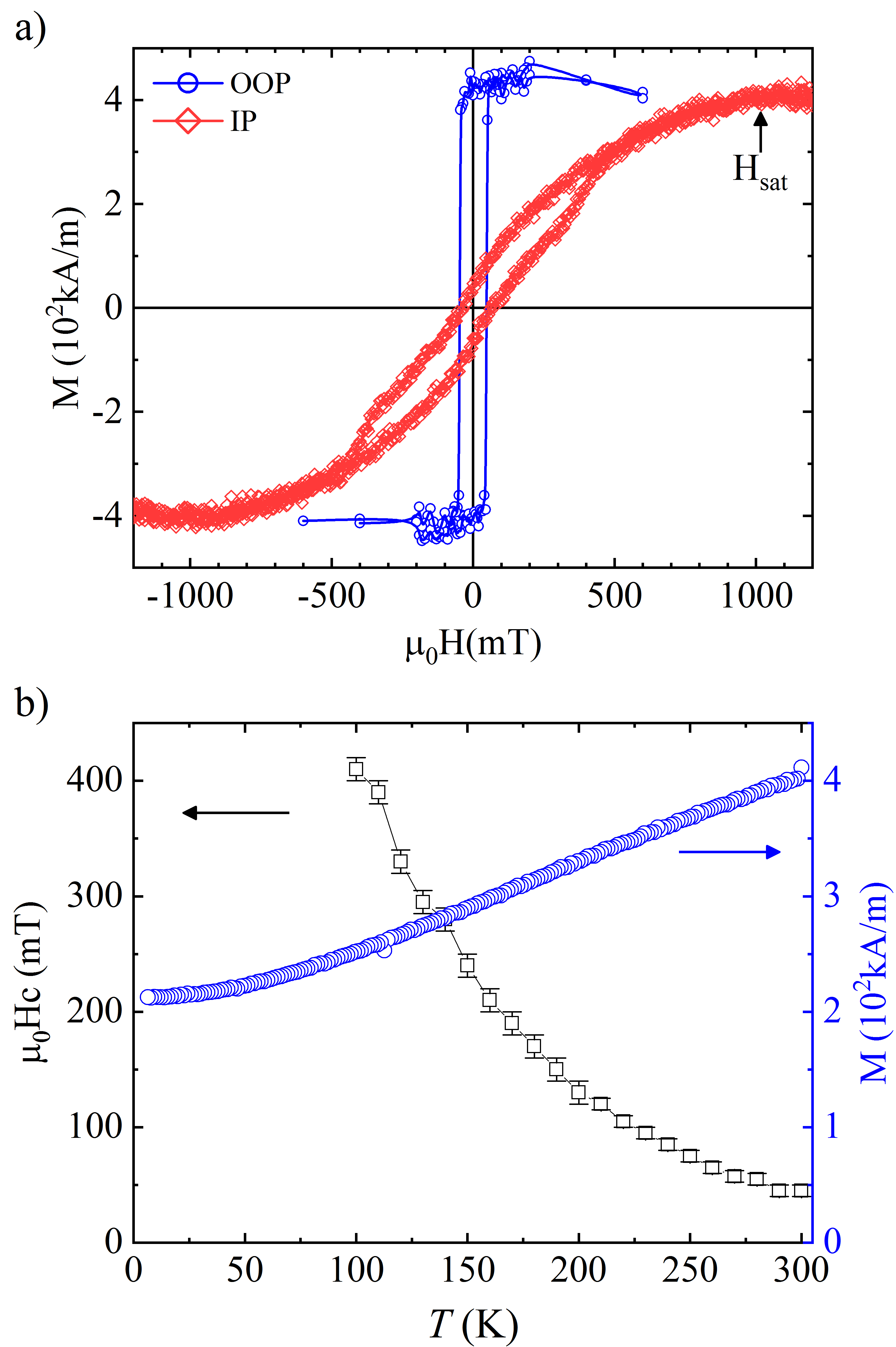}
\caption{\label{fig:MOneSample}
Magnetization behavior corresponding to the TbCo(\qty{0.66}{\nm}) multilayer. (a) Hysteresis loops for the out-of-plane (OOP) and in plane (IP) field sweeps at \qty{300}{\kelvin}). (b) Temperature dependence of the saturation magnetization (blue open circles) and coercive field (black open squares) for the same sample.}
\end{figure}

All TbCo(\xTb) samples are synthetic ferrimagnets composed by rare earth-transition metal (RE-TM) multilayers antiferromagnetically coupled. As other ferrimagnets the total magnetization is given by uncompensated contributions. To investigate the magnetic properties of the samples, we use a superconducting quantum interference device (SQUID) and a vibrating-sample magnetometer (VSM).
Most of the samples present a dominant perpendicular magnetic anisotropy. In Fig. \ref{fig:MOneSample}a) we show the in-plane and out-of-plane hysteresis loops measured at room temperature of sample TbCo(\qty{0.66}{\nm}) and in Fig. \ref{fig:MOneSample}b) we show the temperature dependence of the coercive field (\hc) and saturation magnetization (\ms), measured without applied field after saturation at \qty{300}{\kelvin}. 

As shown in Fig. \ref{fig:MOneSample} a), when the magnetic field is applied perpendicular to the film plane (OOP), a square hysteresis loop, typically observed in systems with strong PMA, is obtained. In contrast, when the field is applied in the film plane (IP), the response corresponds to that of a typical hard axis. From this last measurement we can estimate the anisotropy field (\hsat) as the value at which saturation is reached. As the coercive field is significantly smaller than the anisotropy field, the magnetization reversal proceeds by the nucleation and propagation of domain walls, as usually observed in this kind of systems. \cite{Thorarinsdottir2023, Mishra2023, ErtlJMMM1992}

Furthermore, the TbCo(\qty{0.66}{\nm}) sample measured in OOP configuration, presents square hysteresis loops in the whole range of temperatures from \qty{100}{\kelvin} to \qty{300}{\kelvin}. 
The values of \hc~and \ms~obtained at different temperatures are plotted in Fig. \ref{fig:MOneSample}b), black open squares and blue open circles, respectively.
The \ms~was measured in the remanent state, i.e. after OOP saturation, in order to avoid unwanted contributions.
It is evidenced in Fig. \ref{fig:MOneSample}b), that, when the sample is cooled down, the \ms~decreases and the coercive field increases at the same time.
Such temperature dependence of the \ms~is related to the thermal evolution of the uncompensated moments of the Tb and Co sub-lattices. 
At room temperature, the Co contribution dominates, and the net magnetization aligns with the Co magnetic moments, which is characteristic of Co-rich compositions.
At lower temperatures an increment of magnetic moment of the Tb sub-lattice produces a decrease of the net magnetization. \cite{NawateIEEE1990, ErtlJMMM1992}

\begin{figure}[htpb]
\centering
\includegraphics[width=\columnwidth]{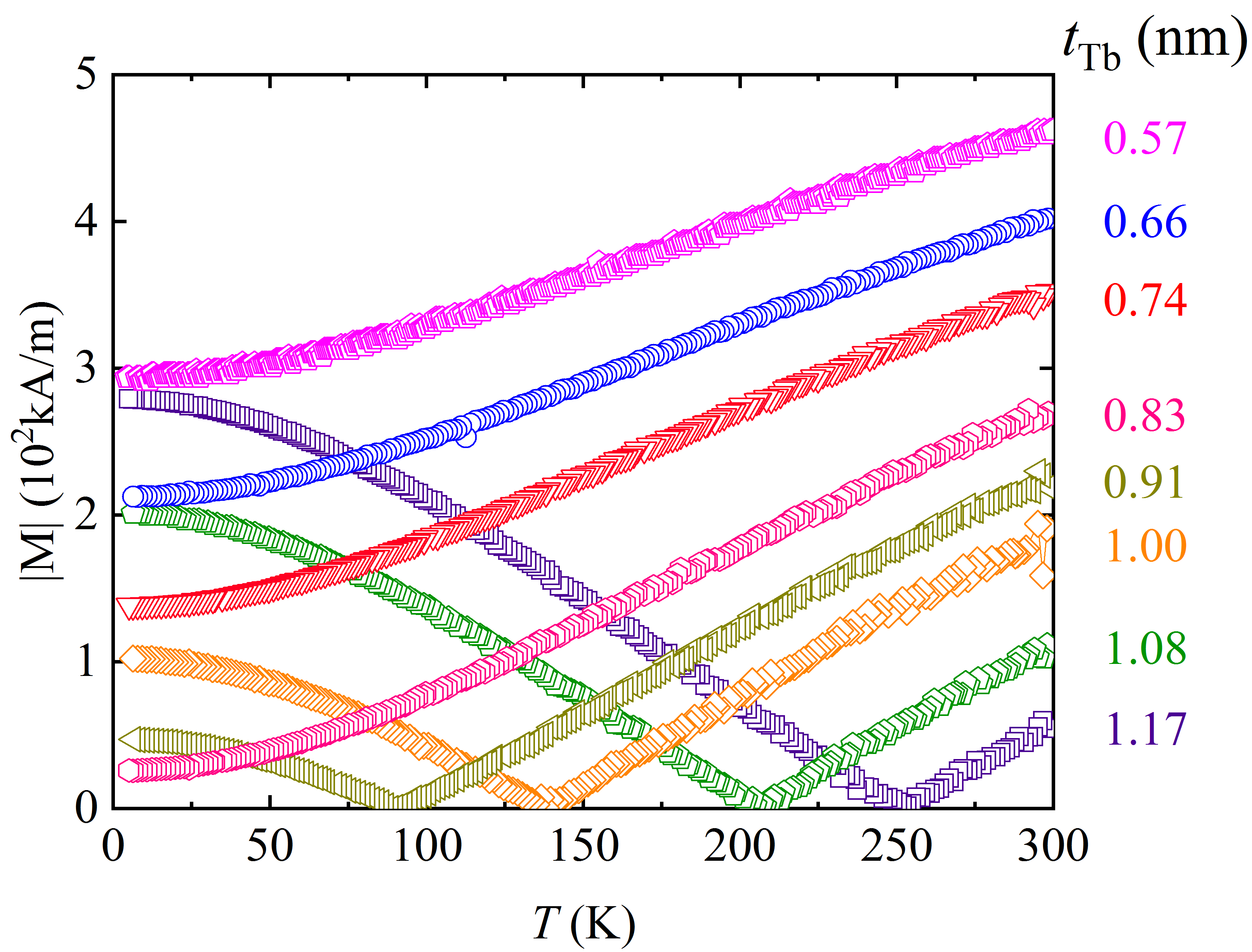}
\caption{\label{fig:MdeT}
Absolute remanent OOP magnetization as a function of temperature for TbCo($t_\mathrm{Tb}$) multilayers with different Tb thicknesses \xTb, as indicated in the labels.
}
\end{figure}

A similar behavior was observed for most samples. In Fig. \ref{fig:MdeT}, the absolute value of the \ms~is shown for different thicknesses \xTb, applying the same protocol described above.
At room temperature, the magnetization of the whole sample set is Co-dominant and decreases when the samples are cooled down.
Samples with \xTb$\leq 0.83$, are Co-dominant in the full temperature range.
For samples TbCo(\xTb) with \xTb$\geq 0.91$ there is a finite temperature (defined as the compensation temperature $\mathrm{T_{M}}$) at which moments of both sub-lattices are compensated resulting in zero net magnetization. Below $\mathrm{T_{M}}$, the magnetization is dominated by the Tb sub-lattice. This property is usually observed in multilayers \cite{RosenkampThesis, NawateIEEE1990} and alloys. \cite{Hansen1989}

\begin{figure}[htbp] 
\centering
\includegraphics[width=\columnwidth]{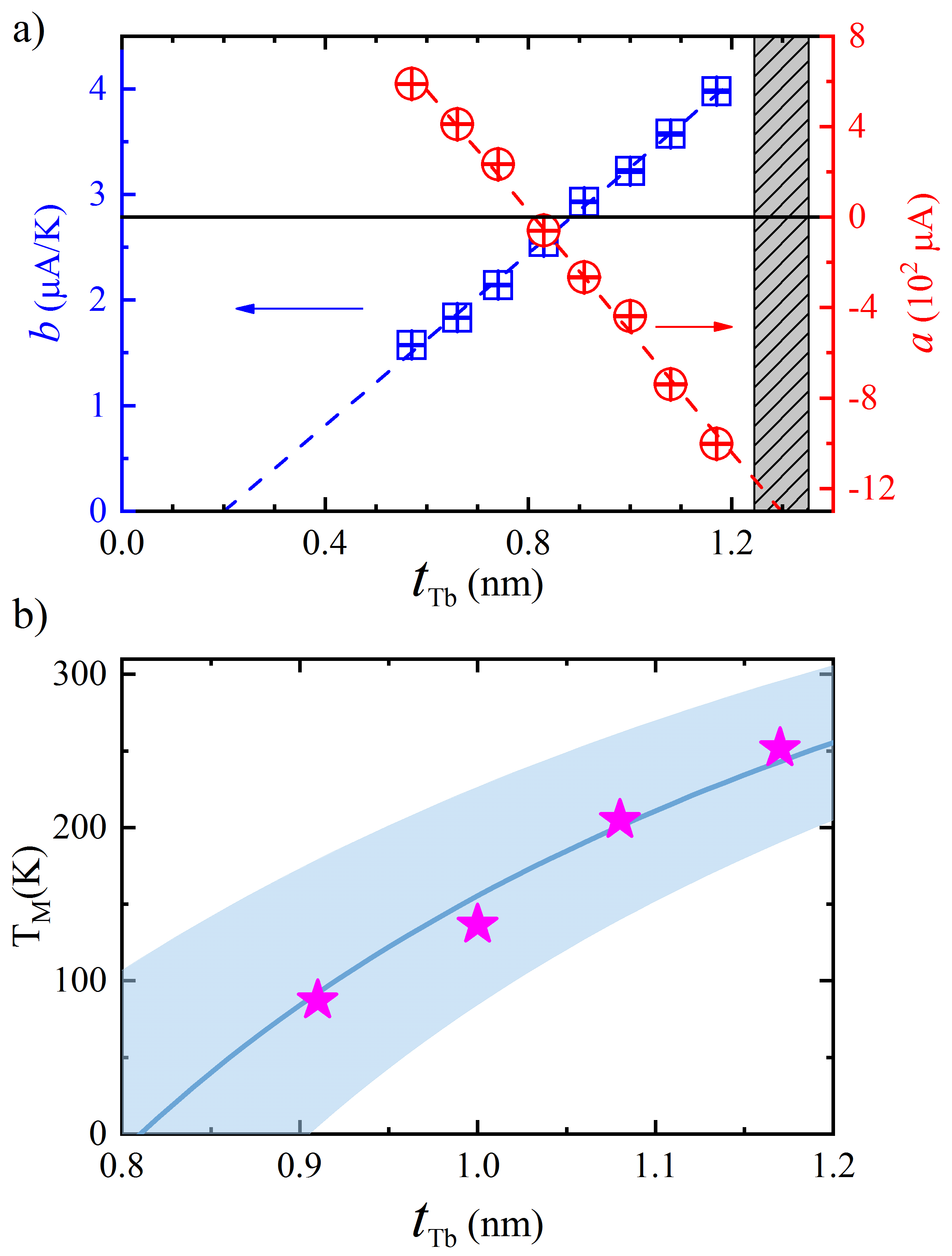}
\caption{\label{fig:ModelTm}
a) Slope $b(t_\mathrm{Tb})$ and intercept $a(t_\mathrm{Tb})$ as a function of \xTb~ extracted from the different measurements shown in Fig. \ref{fig:MdeT} in the range $T \in[\qty{80}{\kelvin}, \qty{280}{\kelvin}]$. The dashed lines correspond to linear fits, as explained in the text. b) Magnetic compensation temperature TM as a function of the tTb. The blue solid line was calculated using the Eq. \eqref{eq:ModelTm}, while the magenta solid stars represent the $\mathrm{T_{M}}$ values extracted from the measurements shown in Fig. \ref{fig:MdeT}. The uncertainty in each $\mathrm{T_{M}}$ determination is   $\qty{+-2}{\kelvin}$, which is much smaller than the symbol size and has therefore been omitted for clarity. The uncertainties associated with the TM computed by Eq. \eqref{eq:ModelTm} is represented by the blue shaded region.
}
\end{figure}

Two remarkable features can be extracted from Fig. \ref{fig:MdeT}. There is a wide temperature range where the slope of the remanent magnetization as a function of temperature is almost constant and, in addition, the value of that slope is very similar across the whole set of samples. Consistently, previous studies have reported that the magnetic moment of the Co sub-lattice remains nearly constant within the explored temperature range \cite{Hu_SperimagneticAmAll_PRB2025, Myers1951, Hu_YCoAlloys_PRB2024}, indicating that the observed temperature dependence originates primarily from the Tb contribution.
However, the observed behavior is also compatible with a weak temperature dependence of the cobalt magnetization, as reported by other authors \cite{LAvilesSR2021}. Therefore, by assuming a linear temperature dependence for the magnetization of both Co and Tb sub-lattices in the range  $T$ = [\qty{80}{\kelvin}, \qty{280}{\kelvin}], we can extract the individual Co and Tb contributions to the total magnetization ($M(T)=M^{\mathrm Co} - M^{\mathrm Tb}$) by fitting the experimental data shown in Fig. \ref{fig:MdeT}.
In this framework, we express the magnetization of each component as $M^{\mathrm{Tb}}(T)=M_S^{\mathrm{Tb}}(1-\alpha T)$ and $M^{\mathrm{Co}}(T)=M_S^{\mathrm{Co}}(1-\alpha_2 T)$, where $M_S^{\mathrm{Tb}}$ and $M_S^{\mathrm{Co}}$ are the spontaneous magnetizations of terbium and cobalt at absolute zero and $\alpha$ and $\alpha_2$ are constant parameters introduced to fit the model to the experimental data.
For clarity, we adopt the following notation for layer thicknesses: $t_\mathrm{Tb}$ for Tb, $t_\mathrm{Co}$ for Co, and $t$ for the total thickness of each bilayer of Tb/Co ($t=t_\mathrm{Tb}+t_\mathrm{Co}$). Accordingly, the total magnetization multiplied by the total thickness of the sample can be written as:
\begin{eqnarray}
t \cdot M(T) &=& M_S^{\mathrm{Co}}(1-\alpha_2 T)t_\mathrm{Co} - M_S^{\mathrm{Tb}}(1-\alpha T))t_\mathrm{Tb}\nonumber\\
&=&a(t_\mathrm{Tb})+b(t_\mathrm{Tb})T,
\label{eq:MTotalT}
 \end{eqnarray}
where the Co thickness ($t_\mathrm{Co}$) is the same for all samples and $a$ and $b$ are functions of the Tb thickness (\xTb) given by:
\begin{eqnarray}
 a(t_\mathrm{Tb}) &=& M_S^{\mathrm{Co}}t_\mathrm{Co}-M_S^{\mathrm{Tb}}t_\mathrm{Tb}, \label{eq:aTb}\\
 b(t_\mathrm{Tb}) &=& -\alpha_2 M_S^{\mathrm{Co}}t_\mathrm{Co} + \alpha M_S^{\mathrm{Tb}}t_\mathrm{Tb}. \label{eq:bTb}
 \end{eqnarray}
Note that the temperature dependence of $M(T)$ is explicitly given by Eq.\eqref{eq:MTotalT}.
In Fig. \ref{fig:ModelTm}a) we plot the obtained $a(t_\mathrm{Tb})$ and $b(t_\mathrm{Tb})$, for the different measurements shown in Fig. \ref{fig:MdeT} in the range $T \in[\qty{80}{\kelvin}, \qty{280}{\kelvin}]$, as a function of the \xTb.
We can clearly see a linear behavior as a function of \xTb~(symbols), together with the linear fits represented by the dashed lines.

The good agreement between Eq. \eqref{eq:aTb} and \eqref{eq:bTb} and the experimental results indicates that our assumptions regarding the magnetization—namely, a linear temperature dependence for both the Tb and Co sub-lattices—are well justified.
Remarkably, one of the quantities extracted from the fit of Eq. \eqref{eq:aTb} is the saturation magnetization of Tb. The obtained value, $\mathrm{M_S^{Tb}}=\qty{26.7+-0.8e2}{\kA/m}$, is very close to the value of  \qty{27.2+-0.3e2}{\kA/m} reported by D. E. Hegland \textit{et al.} \cite{Hegland1963}, as well as to the value used in the simulations of the thermal evolution of the magnetization in Tb/Co multilayers by L. Avilés-Félix \textit{et al.} \cite{LAvilesSR2021}. This agreement further supports the validity of the assumptions employed in our analysis. It is worth noting that the choice of the temperature window was not arbitrary. All measured samples exhibit a clear and robust linear temperature dependence of the total magnetization within the $\qty{80}{\kelvin} - \qty{280}{\kelvin}$ range, as shown in Fig. \ref{fig:MdeT}, which is consistent with the assumption that the parameters $a(t_\mathrm{Tb})$ and $b(t_\mathrm{Tb})$ are temperature independent. Outside this interval, deviations from linearity become apparent. Below approximately $\qty{80}{\kelvin}$, the Tb sublattice magnetization starts to saturate, as also reported in previous studies \cite{LAvilesSR2021, Hegland1963,KudyukovIEEETrMag2025}, while above $\qty{280}{\kelvin}$ the experimental data in Fig. \ref{fig:MdeT} show, in some cases, noticeable departures from linear behavoir. For the reasons, the analysis was restricted to the $\qty{80}{\kelvin} - \qty{280}{\kelvin}$ temperature range and the same window was applied consistently to all samples.

Based on the previous observations we also expect a direct influence of the Tb and Co thicknesses on the magnetic  compensation temperature. As $\mathrm{T_{M}}$ corresponds to the temperature at which the magnetic moments of both sub-lattices are compensated (meaning that Eq.\eqref{eq:MTotalT} must be set to zero). For instance, we can test for samples with $t_{\mathrm{Tb}}\geq 0.91$ if the observed increase of the $\mathrm{T_{M}}$, when \xTb~increases, is purely related to the increment of the total amount of Tb within the sample. Under same assumed hypothesis, we find that the $\mathrm{T_{M}}$ scales with the other quantities as: 
\begin{equation}
   \mathrm{T_{M}} = \frac{1}{\alpha M_S^{\mathrm{Tb}}}\left[M_S^{\mathrm{Tb}} - \frac{M_S^{\mathrm{Co}}t_\mathrm{Co}-M_S^{\mathrm{Tb}}{t^*_\mathrm{Tb}}}{t_\mathrm{Tb}-t^*_\mathrm{Tb}}\right]. \label{eq:ModelTm}
 \end{equation}
Where  $t^*_\mathrm{Tb} =\frac{\alpha_2 M_S^{\mathrm{Co}}}{\alpha M_S^{\mathrm{Tb}}}t_\mathrm{Co}$ is a parameter determined by the saturation magnetization of the two sub-lattices and by the Co layer thickness. The experimental value, $t^*_\mathrm{Tb} = 0.2$ nm, corresponds to the Tb thickness at which the parameter $b$ vanished, i. e., $b(t^*_\mathrm{Tb}) = 0$.
Note that Eq. (\ref{eq:ModelTm}) describes the dependence observed in Fig. \ref{fig:MdeT}, evidencing that $\mathrm{T_{M}}$ increases because the second term is getting smaller as the \xTb~increases.
In Fig. \ref{fig:ModelTm}b) we plot the compensation temperature $\mathrm{T_{M}}$ as a function of \xTb~extracted from Fig. \ref{fig:MdeT} (magenta stars), together with the results of Eq. \eqref{eq:ModelTm} (blue line), evaluated using the mean value of the parameters obtained from fitting the experimental data with Eqs.\eqref{eq:aTb} and \eqref{eq:bTb}.
The uncertainty in each $\mathrm{T_{M}}$ determination is  $\pm2$ K, which is much smaller than the symbol size and has therefore been omitted for clarity. In addition, the uncertainty in the $\mathrm{T_{M}}$ values computed from Eq. \eqref{eq:ModelTm} is shown as the blue shaded region, obtained by propagating all parameter uncertainties extracted from the fits to the experimental data and by conservatively assuming a 1 \AA~error margin in the Tb thickness.
The good agreement between the measured and the estimated values for $\mathrm{T_{M}}$ evidences that the $\mathrm{T_{M}}$ as a function of \xTb~is mainly driven by the Tb/Co ratio of each sample. Indeed, the same trend was reported by other authors for $\mathrm{T_{M}}$ beyond room temperature in multilayers with higher Tb compositions \cite{Alebrand2012, LAvilesSR2021} or below room temperature in Gd alloys. \cite{Nava2025}

\begin{figure}[htbp] 
\centering
\includegraphics[width=\columnwidth]{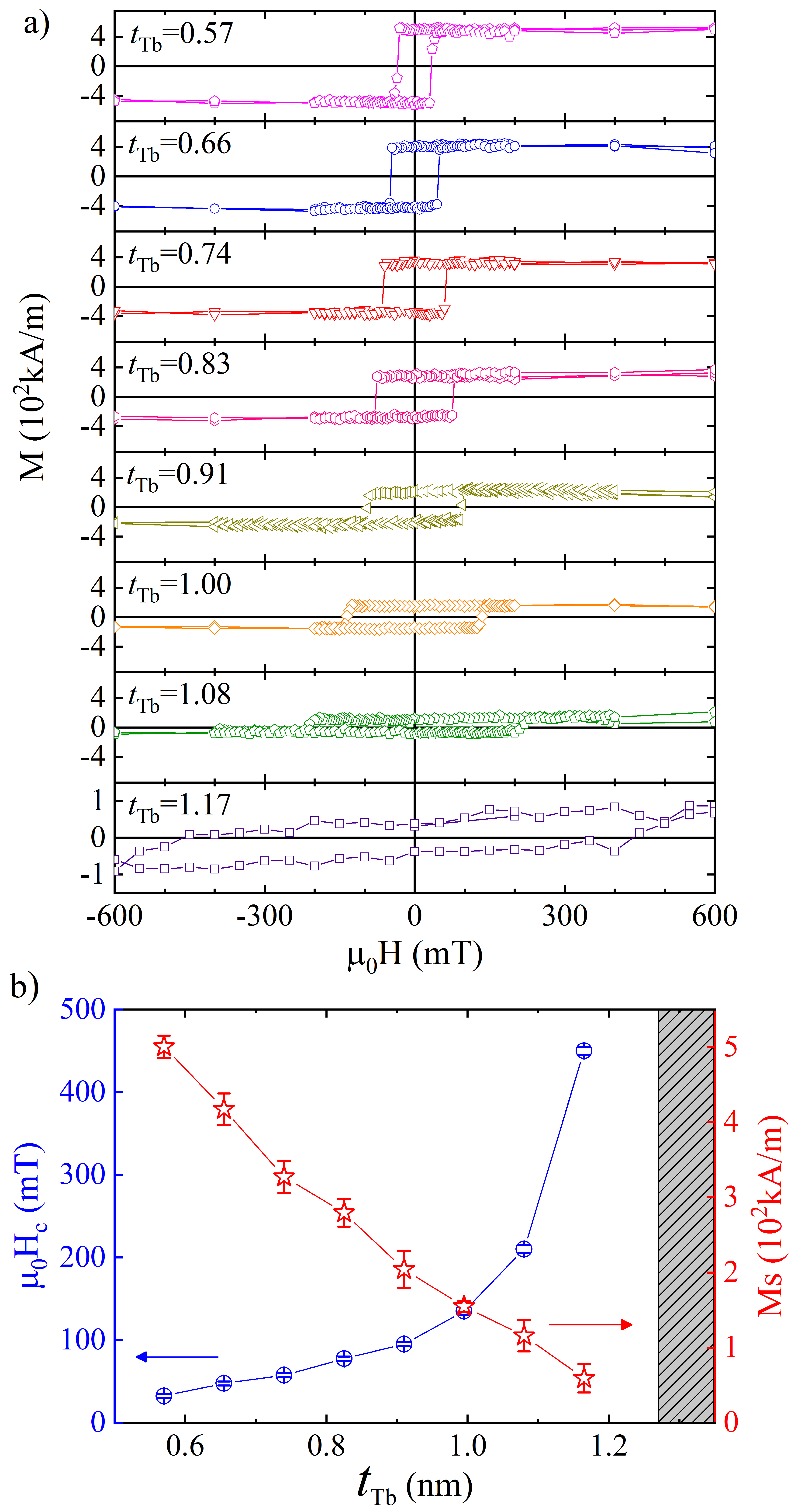}
\caption{\label{fig:MdHAllSamples}
a) Evolution of the OOP magnetic hysteresis loops with the \xTb~at room temperature. b) Coercive field  (blue circles) and saturation magnetization (red stars) dependence on the \xTb~. Grey dashed area corresponds to the thickness of Tb layers $\mathrm{t_{com}}$ in which composition has a magnetization compensation point close to room temperature.
}
\end{figure}

Continuing with the study of the  magnetic properties of the sample set, in Fig. \ref{fig:MdHAllSamples}a), we present the OOP hysteresis cycles at room temperature that were carried out by VSM magnetometry. We can see that all hysteresis cycles are square shaped, evidencing a dominant PMA. Increasing the \xTb, the \hc~increases while \ms~decreases, concomitant with the results shown in Fig. \ref{fig:MdeT} at room temperature. These magnetic properties are summarized in Fig. \ref{fig:MdHAllSamples}b), where it is possible to see that for a \xTb = \qty{1.3}{\nm}, the \ms~is fully compensated and \hc~diverges at room temperature, as usually observed in multilayers \cite{LAvilesSR2020, Frackowiak2022} and alloys of similar composition.\cite{Ciuciulkaite2020, Alebrand2012}

Most of the results described up to now are commonly observed in thin ferrimagnetic films. However one important feature observed in the samples that can be highlighted. As it is possible to see in Fig. \ref{fig:MOneSample} for sample TbCo(\qty{0.66}{\nm}), the IP cycle presents hysteresis with a non negligible \hc.
This fact, unexpected for a uniaxial magnetic systems, was also previously observed by other authors\cite{Mishra2023,ErtlJMMM1992}, and motivated a deeper and more careful study of the magnetic anisotropies, presented in the following subsection.

\subsection{Characterization of the magnetic anisotropies}
\begin{figure}[htbp]
\centering
\includegraphics[width=\columnwidth]{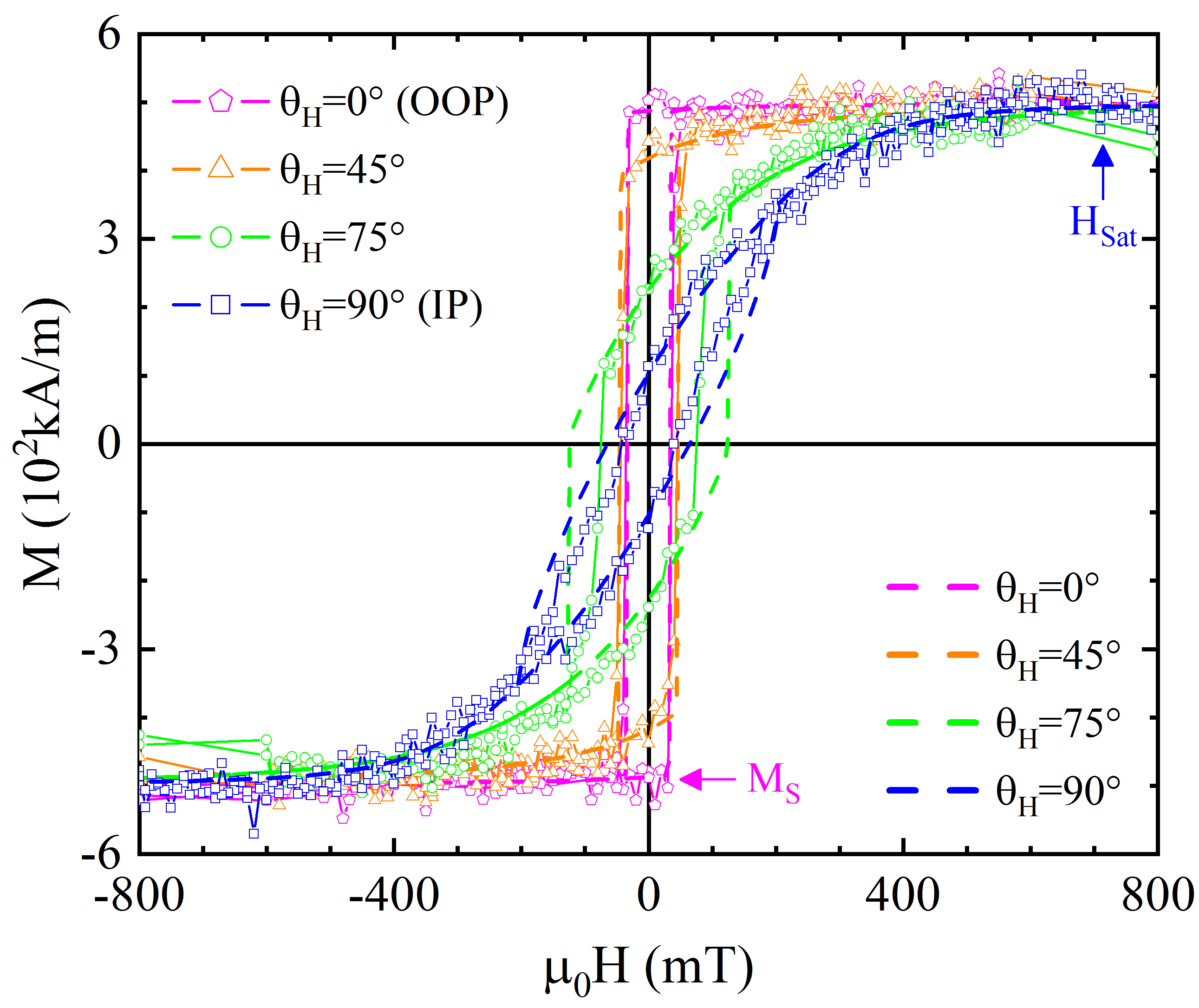}
\caption{\label{fig:MdH057DiffThetas}
Hysteresis loops for TbCo(\qty{0.57}{\nm}) multilayer at different relative orientations of the external magnetic field and the sample plane, from OOP ($\theta_{H} =0^{\circ}$) to IP ($\theta_{H} =90^{\circ}$) at room temperature. The lines and symbols correspond to experimental data, while the dashed lines correspond to simulations.
}
\end{figure}

To better understand the origin of the observed hysteresis in the IP cycles, we performed hysteresis cycle measurements on different samples.
Since the results obtained for different samples are equivalent, for the description of the measurements we will focus the discussion on the results corresponding to a representative sample TbCo(\qty{0.57}{\nm}).
Fig. \ref{fig:MdH057DiffThetas} shows different hysteresis loops acquired for this sample. Here, the polar ($\theta_{H}$) and azimuthal ($\phi_{H}$) angles of the external magnetic field were changed from the OOP configuration ($\theta_{H}=0 ^{\circ}$) to the IP one [$(\theta_{H},\phi_{H})=(90 ^{\circ},0 ^{\circ})$]. The sample is contained within the $\hat x-\hat y$ plane when $(\theta_{H},\phi_{H})=( 90^{\circ}, 0^{\circ})$ which corresponds to $\hat x$ axis and also the direction of the Tb thickness gradient.

The symbols in Fig. \ref{fig:MdH057DiffThetas} correspond to the experimental results measured at 
$\theta_{H} =0 ^{\circ}, 45 ^{\circ},75 ^{\circ}$ and $90 ^{\circ}$, while the dashed lines correspond to the simulations performed at the same angles. It can be observed that the shape and coercive fields of the loops change with the orientation of the external magnetic field, and a non negligible coercive field is observed for all cases.
In the OOP condition ($\theta_{H} =0 ^{\circ}$), a square shaped hysteresis loop is observed. This is usually associated to the typical behavior of the magnetization when the magnetic field is oriented in the direction of an easy axis. In contrast, the IP measurement ($\theta_{H} = 90^{\circ}$), shows different characteristics: in the low field region, the magnetization shows an approximately linear increase with the magnetic field up to the saturation field value \hsat~, which is the same saturation value as that obtained for the OOP measurement. Both characteristics are expected when the magnetic field is oriented close to the perpendicular direction of an easy axis.
Another feature to highlight regarding the hysteresis loops is that by comparing the IP and OOP measurements we can notice that the coercive field obtained for the OOP case is much smaller than the IP field needed to saturate the sample, close to 700 mT. This feature is generally associated with a magnetization reversal process driven by domain nucleation and domain wall propagation rather than a coherent rotation of the magnetization\cite{Barbara1976, Barbara1994}.
Regarding the shape of the hysteresis cycles measured at intermediates polar angles, we can observe in Fig. \ref{fig:MdH057DiffThetas} a smooth evolution of the shape of cycles between the OOP and IP loops.
We also performed hysteresis cycle measurements on the same sample by applying an in plane magnetic field in different orientations $\phi_{H} \in[0^{\circ}, 180^{\circ}]$. For all cases we obtained results that are similar to the IP measurement with $\phi_{H} = 0^{\circ}$ (see blue symbol in Fig. \ref{fig:MdH057DiffThetas}).

As mentioned above, the shape of the OOP hysteresis cycles of all samples suggests a dominant PMA, but at the same time, since we are dealing with films whose \ms~is non negligible (see Fig. \ref{fig:MdHAllSamples}b)), a magnetostatic contribution term is also expected. This contribution, contrary to that of the PMA, induces the magnetization to fall into the plane of the sample.
In addition, if the uniaxial PMA and the shape anisotropy (hard axis in $\hat z$) are both collinear, they can be treated as an effective uniaxial anisotropy and, contrary to the observation, no hysteresis should be present in the IP measurements (see blue symbol in Fig. \ref{fig:MdH057DiffThetas}). 
Even if we assume that due to the Tb thickness gradient there is an extra uniaxial anisotropy in the sample plane (along $\hat x$), the result indicates that hysteresis in the IP measurements should \textit{not} be observed. 
To better understand the role of anisotropies and their impact on the magnetic properties of the samples, we further explored modeling to qualitatively and quantitatively reproduce the experimental results. As we will discuss in the following section, a mandatory condition that the system must satisfy in order to observe hysteresis in the IP cycles is that some of the uniaxial anisotropies must be at least slightly tilted off-axis. 

\subsection{Modeling and simulations}

In the previous subsections we presented and discussed magnetization measurements for different Tb thicknesses as a function of temperature and magnetic field for different orientations. By comparing the IP and OOP measurements we note that the magnetization reversal process is driven by domain nucleation and domain wall propagation. However, at the time scale of the measurement of the hysteresis loop, the sharp square shape of the OOP loop not only indicates we have dominant uniaxial anisotropy pointing close to the $\hat z$ axis, but also suggests that the reversal of the magnetization is abrupt and behaves as a single magnetic entity.
Furthermore, due to the non-negligible value observed for \ms~at room temperature (see Fig. \ref{fig:MdHAllSamples} b)), the shape anisotropy can not be neglected.
Moreover, there is no reason to suspect that cubic or another non-uniaxial  anisotropy might be present, mainly because the samples are non-crystalline, but also because the measured hysteresis loops do not have the correct symmetry to fit such anisotropies. 

In this scenario, one of the unexpected characteristics of the measurements is the presence of hysteresis in the IP field measurements. This feature, observed for different angles within the interval $\phi_{H} \in[0^{\circ}, 180^{\circ}]$, cannot be explained if PMA and shape anisotropy were uniaxial and collinear or orthogonal.
However, even if the Tb layer thickness can be considered constant across the measured samples, the off-axis growth required to induce the wedge-shaped Tb layers is the only ingredient that breaks the azimuthal symmetry and could therefore be at the origin of this effect. It is well established that placing the substrate in an off-axis configuration affects film growth differently compared to the on-axis geometry. Some of the most relevant differences are related to the irradiation received by the substrate and to the surface mobility of the adatoms, which can in turn have direct consequences for the microstructure and magnetic properties of the multilayer. In particular, a reduced generation of point defects and lower damage-induced roughness are expected. In addition, this symmetry breaking may lead to the development of anisotropic residual stresses, which could be responsible for the observed magnetic anisotropy \cite{EshoAIP1976,JohnsonRPP1996,BubendorffEPL2006,Alfonso2012}.

Thus, taking into account all these considerations, we developed a simplified model to reproduce the experimental results.
In this case, we assumed that the magnetic behavior of the system can be described as a single macrospin ($\overrightarrow{M} =  M_s \hat{m}$) and consequently, it will have only 2 degrees of freedom to define the orientation of the magnetization vector, i.e. $\theta_{m}$ and $\phi_{m}$.
Also, due to the sample dimensions, to compute the demagnetizing factor we assume the shape of the sample as an infinite plane.
Under the above assumptions, the magnetic energy density of the system can be written as:
\begin{equation}
E= -\mu_{0}\overrightarrow{M}\cdot\overrightarrow{H} + \frac{1}{2} \mu_{0} M_s^2 (\hat{m}\cdot \hat{z})^2 - \ku(\hat{m}\cdot \hat{\eta})^2.
\label{eq:EnmagSI}
\end{equation}
Where $E$ is the energy per unit volume. The first term in Eq.\eqref{eq:EnmagSI} corresponds to the Zeeman energy of the system, which accounts for the effect of the external field. The second term corresponds to the shape anisotropy of the system, and the third term corresponds to a uniaxial effective anisotropy that competes with the shape term and confers to the system the dominant PMA behavior.
We would like to emphasize that Eq.\eqref{eq:EnmagSI}, where only two uniaxial anisotropies are present, represents the most simple model that can be proposed for this system. In this scenario, it is important to note that the unit vector $\hat{\eta}$, that defines the direction of the easy axis of the anisotropy, can be oriented close to, but not exactly along the $\hat{z}$ direction. If $\hat{\eta}$ and $\hat{z}$ are coincident, the expression of the anisotropy energies represented by Eq.\eqref{eq:EnmagSI} can be rewritten as a single effective anisotropy term, which cannot reproduce the hysteresis observed in the IP cycle (see blue symbol in Fig. \ref{fig:MdH057DiffThetas}).

To minimize the total energy of the system we implemented an algorithm that combines two methods: Steepest-Descent Algorithm when magnetization is reversed or far from the desired local minimum; and a Newton-like method when the energy of the system is close to the local minimum.
By optimizing the minimization procedure we were able to identify two situations that require essentially different solutions. If the external field is applied in directions far from the normal to the sample plane, the magnetization reversal is determined by the anisotropy of the system. Otherwise, when the external field is applied in directions close to the normal to the sample plane, where nucleation and propagation of domain walls dominate, the magnetization reversal is treated \textit{ad hoc} by suddenly reversing the magnetization. 
In this case, the reversal condition is given by $\overrightarrow{H}\cdot\hat{n}=-H_{nuc}$, where $\hat{n}$ is normal to the sample plane and $H_{nuc}$ is the value of the experimental coercive field in the OOP configuration ($\theta_{H}=0^{\circ}$).
Magnetization reversal is handled by introducing changes $\theta_{m}\longrightarrow 180^{\circ}-\theta_{m}$ and $\phi_{m}\longrightarrow \phi_{m}+180^{\circ}$. To find the best value of the anisotropy constant $\textrm{K}_\textrm{u}$, that best fits \emph{simultaneously} all the hysteresis cycles for different angles (see Fig. \ref{fig:MdH057DiffThetas}), the following procedure was carried out: First, we estimate an effective anisotropy \keff~from the saturation field observed in the IP loop using the well-known expression $\mu_0 \hsat = \frac{2\keff}{\ms}$. Next, the model described by Eq.\eqref{eq:EnmagSI} turns into the uniaxial anisotropy when $\hat{\eta}$ and $\hat{z}$ are considered as collinear, and therefore the uniaxial anisotropy $\mathrm{K_{u}^{\hat{z}}}$ can be estimated by the following relation $\keff = \mathrm{K_{u}^{\hat{z}}} - \frac{\mu_0 \ms^2}{2}$. It is important to note that this value for $\mathrm{K_{u}^{\hat{z}}}$ is used just as a seed value for our model. 
From Fig. \ref{fig:MdH057DiffThetas}, for the multilayer TbCo(\qty{0.57}{\nm}), the saturation field was estimated as $\mu_0 \hsat = 700 \pm 100$ mT and the  saturation magnetization $\mathrm{M_S}=\qty{500+-50}{\kA/m}$. These quantities give an initial value for $\mathrm{K_{u}^{\hat{z}}}= \qty{330+-30}{\unit{\kilo\J}/\unit{\m}^3}$ that is close to what was reported by Alebrand and coworkers \cite{Alebrand2012} in similar compositions.\footnote{Same composition corresponds to multilayers where the normalized amount of each metal is equivalent to the alloy's atomic fraction $\mathrm{Tb}_{x}\mathrm{Co}_{1-x}$} Finally, we iteratively optimized $\mathrm{K_{u}}$ and the tilt angle $\hat{\eta}$ with respect to its initial configuration ($\hat{\eta}=\hat{z}$), thereby increasing the in-plane coercive field and reproducing the experimentally observed loop shapes.

The best results were obtained when the deviation of $\hat{\eta}$ from the $\hat{z}$ axis was $4^{\circ}$ into the $\hat{z}$-$\hat{x}$ plane and for \mbox{$\ku = \qty{230+-30}{\unit{\kilo\J}/\unit{\m}^3}$}. It is important to mention that it is impossible to reproduce the observed hysteresis loops in the in-plane configuration by simply tilting a single uniaxial effective anisotropy away from the  $\hat{z}$ axis. In this context, we emphasize that such anisotropies are effective anisotropies, and therefore do not necessarily coincide with the local microscopic anisotropies that give rise to sperimagnetic or related non-collinear behaviors in RE–TM alloys or multilayers. Moreover, the presence of a tilted effective anisotropy axis also implies that the perpendicular magnetic anisotropy and the in-plane shape anisotropy can no longer be directly combined into a single effective uniaxial anisotropy for the sample, since the latter remains a hard axis strictly aligned with the film normal. Therefore, our samples must be considered as systems exhibiting two distinct uniaxial anisotropies. The results of our simulations are shown in Fig. \ref{fig:MdH057DiffThetas} as dashed lines of the same color as the corresponding experimental data. As can be seen, the simulations of the hysteresis loops are in good agreement with the measured data.

In addition, an analogous analysis can be carried out for the sample TbCo(\qty{0.66}{\nm}), where we found a similar values for the anisotropy $\ku \approx \qty{230}{{\kilo\J}/{\m}^3}$ and tilting $\hat{\eta}= 4^{\circ}$, as for $t_{\mathrm{Tb}} = 0.57$. These values are also compatible with the $t_{\mathrm{Tb}} = 0.74$ and $0.83$ samples, however, in these cases new features emerge in the hysteresis loops, indicating that the model progressively fails to capture the system’s behavior. This deviation becomes increasingly pronounced as the Tb thickness increases, pointing to differences in the reversal process. Although studying these features could provide valuable insight into the design and optimization of Tb/Co multilayers for applications, a detailed analysis lies beyond the scope of the present work.
For samples with $t_{\mathrm{Tb}}\geq 0.91$, it was no longer possible to determine \hsat~since it exceeds the maximum field available.

Notably, Salomoni \textit{et al.} \cite{SalomoniAPS2023} also conclude that tilting the uniaxial anisotropy axis away from $\hat{z}$ is essential to explain the emergence of reversal-ring patterns observed in all-optical switching experiments. Typical tilt angles of $5^\circ$ were needed to produce a good qualitative agreement between simulations and experimental observations.

\section{Summary and Conclusions}

To conclude, we report the fabrication and magnetic characterization of multilayers composed of five periods Tb/Co with graded thicknesses of the Tb layer and constant thicknesses of Co layer. The structure and composition of the samples was confirmed by HAADF-STEM, EDX and PIXE techniques. From the experimental results it was evidenced that the relative fraction of Tb and Co is very close to its nominal values and the interfaces are well defined with non-negligible interdiffusion effects.

We investigated the influence of Tb on the magnetic properties of these Tb/Co multilayers with varying Tb thickness and temperature. The magnetic characterization shows a clear Tb thickness dependence of the magnetic properties. All the samples exhibited a strong perpendicular magnetic anisotropy that decreases with increasing Tb layer thickness. We reported that the magnetization compensation point at room temperature corresponds to a Tb layer thickness of $\qty{1.3}{\nm}$. In addition, the temperature dependence of the remanent magnetization indicates that both Co and Tb magnetizations vary approximately linearly with temperature, with the temperature dependence being substantially weaker for Co. Under this hypothesis, the relation of the compensation temperature $\mathrm{T_{M}}$ with the Tb thickness was successfully explained by assuming that the net magnetization of the system is the sum of both contributions. The observed scale with the Tb thickness confirms the good quality of the Co and Tb layers as well as the fine control of the Tb thickness gradient.

Measurements of the hysteresis loops as a function of external field orientation show an evolution of their shape with angle. The comparison between the values measured from \hsat~and \hc~indicates that the magnetization reversal process is driven by the nucleation and propagation of domain walls. Hence, it was necessary to consider the magnetization reversal by domain nucleation. We demonstrated that our simplified model reproduces qualitatively and quantitatively the shape of the hysteresis loops with only two anisotropies: one of them is the shape anisotropy which points in $\hat z$ axis, hard axis, and the other is a slightly tilted uniaxial anisotropy from $\hat z$, $4^\circ$ in our case. Even though the effective uniaxial anisotropy value obtained here is in good agreement with those reported by other authors\cite{Alebrand2012, SalomoniAPS2023}, we show that the extracted value, $\ku \approx \qty{230}{{\kilo\J}/{\m}^3}$, is not directly comparable to anisotropy values reported in the literature that assume a strictly $\hat z$-orieted uniaxial anisotropy ($\mathrm{K_{u}^{\hat{z}}}$). Subsequently, we explained the hysteresis observed for in-plane field hysteresis cycles. The microscopic origin of the experimentally observed tilting of the effective easy axis has not yet been conclusively identified. However, a plausible contribution may be associated with the growth conditions of the samples, in particular with the asymmetry of the sputtering geometry that gives rise to a slight Tb thickness gradient along the sample. Although this gradient is very small, the resulting anisotropic irradiation may provide a microscopic mechanism—such as the development of anisotropic residual stresses—that could be responsible for the observed tilting of the magnetic anisotropy. Further investigation is required to establish and characterize the microscopic origin of this phenomenon. This work contributes to understanding the DC magnetic properties of this kind of ferrimagnetic multilayers. These properties are crucial in areas such as magnetotransport or all-optical-switching , where the anisotropy plays a key role in the magnetization relaxation.

\section{Acknowledgments}
 The authors would like to thank H. Pastoriza and L. Salazar Alarcón from Centro Atómico Bariloche for their technical assistance. We also thank EU-H2020-RISE project Ultra Thin Magneto Thermal Sensoring ULTIMATE-I (Grant ID. 101007825), Magnetism and Cryogenics Competence Center, and specially to C. Rojas-Sánchez  and M. Yactayo from Institute Jean Lamour for complementary hysteresis loops measurements on the samples. This work was partly supported by the Argentine MinCyT Projects \mbox{PICT-2021-I-INVI-00113} (project DISCO) and PICT 2019-2873, and by the University of Cuyo by Grants No. 06/C035-T1, 06/80020240100271UN and 06/80020240100273UN, and by Conicet under Grant PIBAA 2022-2023 (project MAGNETS) Grant ID. 28720210100099CO.


\bibliographystyle{elsarticle-num}

\bibliography{TbCoMagnetic}

@article{KudyukovIEEETrMag2025,
	author = {Kudyukov, Egor V. and Nizaev, Azat N. and Kravtsov, Evgeniy A. and Gorkovenko, Alexander N. and Lepalovskij, Vladimir N. and Vas'Kovskiy, Vladimir O.},
	doi = {10.1109/TMAG.2025.3573100},
	journal = {IEEE Transactions on Magnetics},
	number = {7},
	pages = {1-9},
	title = {Features of the Asperomagnetic Ordering of Thin Films of Heavy Rare-Earth Metals and Their Alloys With Co},
	volume = {61},
	year = {2025}}

@article{EshoAIP1976,
	author = {Esho, Sotaro and Fujiwara, Shozo},
	doi = {10.1063/1.2946119},
	eprint = {https://pubs.aip.org/aip/acp/article-pdf/34/1/331/11733306/331_1_online.pdf},
	issn = {0094-243X},
	journal = {AIP Conference Proceedings},
	month = {09},
	number = {1},
	pages = {331-333},
	title = {Growth Induced Anisotropy in Sputtered $\mathrm{GdCo}$ Films},
	url = {https://doi.org/10.1063/1.2946119},
	volume = {34},
	year = {1976}}

@article{JohnsonRPP1996,
	author = {M T Johnson and P J H Bloemen and F J A den Broeder and J J de Vries},
	date = {1996/11/01},
	doi = {10.1088/0034-4885/59/11/002},
	isbn = {0034-4885},
	journal = {Reports on Progress in Physics},
	number = {11},
	pages = {1409},
	title = {Magnetic anisotropy in metallic multilayers},
	url = {https://doi.org/10.1088/0034-4885/59/11/002},
	volume = {59},
	year = {1996}}

@article{BubendorffEPL2006,
	author = {J. L. Bubendorff and S. Zabrocki and G. Garreau and S. Hajjar and R. Jaafar and D. Berling and A. Mehdaoui and C. Pirri and G. Gewinner},
	date = {2006/06/02},
	doi = {10.1209/epl/i2006-10089-5},
	isbn = {0295-5075},
	journal = {Europhysics Letters},
	number = {1},
	pages = {119},
	title = {Origin of the magnetic anisotropy in ferromagnetic layers deposited at oblique incidence},
	url = {https://doi.org/10.1209/epl/i2006-10089-5},
	volume = {75},
	year = {2006}}

@incollection{Alfonso2012,
	address = {London},
	author = {Edgar Alfonso and Jairo Olaya and Gloria Cubillos},
	booktitle = {Crystallization - Science and Technology},
	chapter = {15},
	doi = {10.5772/35844},
	editor = {Marcello Rubens Barsi Andreeta},
	publisher = {IntechOpen},
	title = {Thin Film Growth Through Sputtering Technique and Its Applications},
	url = {https://doi.org/10.5772/35844},
	year = {2012}}

@article{Hu_YCoAlloys_PRB2024,
	author = {Hu, Zexiang and Besbas, Jean and Siewierska, Katarzyna and Smith, Ross and Stamenov, Plamen and Coey, J. M. D.},
	doi = {10.1103/PhysRevB.109.014409},
	issue = {1},
	journal = {Phys. Rev. B},
	month = {Jan},
	numpages = {9},
	pages = {014409},
	publisher = {American Physical Society},
	title = {Magnetism, transport, and atomic structure of amorphous binary $\mathrm{Y}_{x}\mathrm{Co}_{1-x}$ alloys},
	url = {https://link.aps.org/doi/10.1103/PhysRevB.109.014409},
	volume = {109},
	year = {2024}}

@article{DrovosekovTwistedM_Multilayers_JMMM2019,
	author = {A.B. Drovosekov and A.O. Savitsky and D.I. Kholin and N.M. Kreines and V.V. Proglyado and M.V. Makarova and E.A. Kravtsov and V.V. Ustinov},
	doi = {https://doi.org/10.1016/j.jmmm.2018.12.022},
	issn = {0304-8853},
	journal = {Journal of Magnetism and Magnetic Materials},
	pages = {668-674},
	title = {Twisted magnetization states and inhomogeneous resonance modes in a $\mathrm{Fe/Gd}$ ferrimagnetic multilayer},
	url = {https://www.sciencedirect.com/science/article/pii/S0304885318326842},
	volume = {475},
	year = {2019}}

@article{Hu_SperimagneticAmAll_PRB2025,
	author = {Hu, Zexiang and Jha, Ajay and Siewierska, Katarzyna and Lenne, Simon and le Berre, Pierre and Dempsey, Nora M. and Stamenov, Plamen and Rode, Karsten and Coey, J. M. D.},
	doi = {10.1103/yzfd-jqvx},
	issue = {1},
	journal = {Phys. Rev. B},
	month = {Jul},
	numpages = {13},
	pages = {014441},
	publisher = {American Physical Society},
	title = {Magnetism of sperimagnetic amorphous \emph{R}$\mathrm{Co}_{3}$ thin films with \emph{R} = $\mathrm{Dy}$, $\mathrm{Tb}$, and $\mathrm{Tm}$},
	url = {https://link.aps.org/doi/10.1103/yzfd-jqvx},
	volume = {112},
	year = {2025}}

@article{SalomoniAPS2023,
	author = {Salomoni, D. and Peng, Y. and Farcis, L. and Auffret, S. and Hehn, M. and Malinowski, G. and Mangin, S. and Dieny, B. and Buda-Prejbeanu, L.D. and Sousa, R.C. and Prejbeanu, I.L.},
	doi = {10.1103/PhysRevApplied.20.034070},
	issue = {3},
	journal = {Phys. Rev. Appl.},
	month = {Sep},
	numpages = {8},
	pages = {034070},
	publisher = {American Physical Society},
	title = {Field-Free All-Optical Switching and Electrical Readout of $\mathrm{Tb}$/$\mathrm{Co}$-Based Magnetic Tunnel Junctions},
	url = {https://link.aps.org/doi/10.1103/PhysRevApplied.20.034070},
	volume = {20},
	year = {2023}}

@article{Barbara1994,
	author = {Bernard Barbara},
	issn = {0304-8853},
	journal = {Journal of Magnetism and Magnetic Materials},
	number = {1},
	pages = {79-86},
	title = {Magnetization processes in high anisotropy systems},
	volume = {129},
	year = {1994}}

@article{Barbara1976,
	author = {Barbara, B and Uehara, M},
	journal = {IEEE Transactions on Magnetics},
	number = {6},
	pages = {997--999},
	publisher = {IEEE},
	title = {Anisotropy and coercivity in SmCo 5-based compounds},
	volume = {12},
	year = {1976}}

@article{Hegland1963,
	author = {Hegland, Donald Eugene and Legvold, S and Spedding, FH},
	journal = {Physical Review},
	number = {1},
	pages = {158},
	publisher = {APS},
	title = {Magnetization and electrical resistivity of terbium single crystals},
	volume = {131},
	year = {1963}}

@article{Myers1951,
	author = {Myers, HP and Sucksmith, Willie},
	journal = {Proceedings of the Royal Society of London. Series A. Mathematical and Physical Sciences},
	number = {1091},
	pages = {427--446},
	publisher = {The Royal Society London},
	title = {The spontaneous magnetization of cobalt},
	volume = {207},
	year = {1951}}

@mastersthesis{RosenkampThesis,
	author = {Rosenkamp, Roy H.},
	journal = {MSc. Thesis, Eindhoven University of Technology},
	month = {Feb.},
	school = {Applied Physics and Science Education. Eindhoven University of Technology (TU/e).},
	title = {All-optical switching in synthetic ferrimagnetic multilayers with terbium},
	year = {2022}}

@article{NawateIEEE1990,
	author = {Nawate, M. and Morikawa, T. and Tsunashima, S. and Uchiyama, S.},
	journal = {IEEE Transactions on Magnetics},
	number = {5},
	pages = {2739-2741},
	title = {Magnetization property of $\mathrm{Tb/Co}$ multilayered films},
	volume = {26},
	year = {1990}}

@article{Thorarinsdottir2023,
	author = {Th{\'o}rarinsd{\'o}ttir, K A and Thorbjarnard{\'o}ttir, B R and Arnalds, U B and Magnus, F},
	date = {2023/03/16},
	doi = {10.1088/1361-648X/acc226},
	isbn = {0953-8984},
	journal = {Journal of Physics: Condensed Matter},
	number = {20},
	pages = {205802},
	publisher = {IOP Publishing},
	title = {Competing interface and bulk anisotropies in $\mathrm{Co}$-rich $\mathrm{TbCo}$ amorphous thin films},
	url = {https://dx.doi.org/10.1088/1361-648X/acc226},
	volume = {35},
	year = {2023}}

@article{Campbell2010,
	author = {J.L. Campbell and N.I. Boyd and N. Grassi and P. Bonnick and J.A. Maxwell},
	issn = {0168-583X},
	journal = {Nuclear Instruments and Methods in Physics Research Section B: Beam Interactions with Materials and Atoms},
	number = {20},
	pages = {3356-3363},
	title = {The Guelph PIXE software package IV},
	volume = {268},
	year = {2010}}

@article{Peng2023,
	author = {Peng, Y and Salomoni, D and Malinowski, G and Zhang, W and Hohlfeld, Julius and Buda-Prejbeanu, LD and Gorchon, J and Verg{\`e}s, M and Lin, JX and Lacour, D and others},
	journal = {Nature Communications},
	number = {1},
	pages = {5000},
	publisher = {Nature Publishing Group UK London},
	title = {In-plane reorientation induced single laser pulse magnetization reversal},
	volume = {14},
	year = {2023}}

@article{Mishra2023,
	author = {Mishra, K. and Blank, T. G. H. and Davies, C. S. and Avil\'es-F\'elix, L. and Salomoni, D. and Buda-Prejbeanu, L. D. and Sousa, R. C. and Prejbeanu, I. L. and Koopmans, B. and Rasing, Th. and Kimel, A. V. and Kirilyuk, A.},
	issue = {2},
	journal = {Phys. Rev. Res.},
	month = {Jun},
	numpages = {13},
	pages = {023163},
	publisher = {American Physical Society},
	title = {Dynamics of all-optical single-shot switching of magnetization in $\mathrm{Tb}$/$\mathrm{Co}$ multilayers},
	volume = {5},
	year = {2023}}

@article{Gottwald2012,
	author = {Gottwald, M. and Hehn, M. and Montaigne, F. and Lacour, D. and Lengaigne, G. and Suire, S. and Mangin, S.},
	issn = {0021-8979},
	journal = {Journal of Applied Physics},
	month = {04},
	number = {8},
	pages = {083904},
	title = {Magnetoresistive effects in perpendicularly magnetized Tb-Co alloy based thin films and spin valves},
	volume = {111},
	year = {2012}}

@article{ErtlJMMM1992,
	author = {L Ertl and G Endl and H Hoffmann},
	issn = {0304-8853},
	journal = {Journal of Magnetism and Magnetic Materials},
	number = {1},
	pages = {227-237},
	title = {Structure and magnetic properties of sputtered $\mathrm{Tb}$/$\mathrm{Co}$ multilayers},
	volume = {113},
	year = {1992}}

@article{Nava2025,
	author = {Nava Antonio, Guillermo and Remy, Quentin and Lin, Jun-Xiao and Le Guen, Yann and Hamara, Dominik and Compton-Stewart, Jude and Barker, Joseph and Hauet, Thomas and Hehn, Michel and Mangin, St{\~A}{\copyright}phane and Ciccarelli, Chiara},
	journal = {Advanced Optical Materials},
	pages = {2500056},
	title = {Origin of the Laser-Induced Picosecond Spin Current Across Magnetization Compensation in Ferrimagnetic $\mathrm{Gd}\mathrm{Co}$},
	year = {2025}}

@article{LAvilesSR2021,
	author = {Avil{\'e}s-F{\'e}lix, L. and Farcis, L. and Jin, Z. and {\'A}lvaro-G{\'o}mez, L. and Li, G. and Yamada, K. T. and Kirilyuk, A. and Kimel, A. V. and Rasing, Th. and Dieny, B. and Sousa, R. C. and Prejbeanu, I. L. and Buda-Prejbeanu, L. D.},
	date = {2021/03/22},
	doi = {10.1038/s41598-021-86065-w},
	id = {Avil{\'e}s-F{\'e}lix2021},
	isbn = {2045-2322},
	journal = {Scientific Reports},
	number = {1},
	pages = {6576},
	title = {All-optical spin switching probability in [$\mathrm{Tb}$/$\mathrm{Co}$] multilayers},
	url = {https://doi.org/10.1038/s41598-021-86065-w},
	volume = {11},
	year = {2021}}

@article{Alebrand2012,
	author = {Alebrand, Sabine and Gottwald, Matthias and Hehn, Michel and Steil, Daniel and Cinchetti, Mirko and Lacour, Daniel and Fullerton, Eric E. and Aeschlimann, Martin and Mangin, St{\~A}{\copyright}phane},
	journal = {Applied Physics Letters},
	month = {10},
	number = {16},
	pages = {162408},
	title = {{Light-induced magnetization reversal of high-anisotropy TbCo alloy films}},
	volume = {101},
	year = {2012}}

@article{Frackowiak2022,
	author = {{\L}ukasz Frackowiak and Feliks Stobiecki and Maciej Urbaniak and Micha{\l} Matczak and Gabriel David Chaves-O`Flynn and Miko{\l}aj Bilski and Andreas Glenz and Piotr Kuswik},
	issn = {0304-8853},
	journal = {Journal of Magnetism and Magnetic Materials},
	pages = {168682},
	title = {Magnetic properties of $\mathrm{Co}$-$\mathrm{Tb}$ alloy films and $\mathrm{Tb}$/$\mathrm{Co}$ multilayers as a function of concentration and thickness},
	volume = {544},
	year = {2022}}

@article{Ciuciulkaite2020,
	author = {Ciuciulkaite, Agne and Mishra, Kshiti and Moro, Marcos V. and Chioar, Ioan-Augustin and Rowan-Robinson, Richard M. and Parchenko, Sergii and Kleibert, Armin and Lindgren, Bengt and Andersson, Gabriella and Davies, Carl S. and Kimel, Alexey and Berritta, Marco and Oppeneer, Peter M. and Kirilyuk, Andrei and Kapaklis, Vassilios},
	issue = {10},
	journal = {Phys. Rev. Mater.},
	month = {Oct},
	numpages = {11},
	pages = {104418},
	publisher = {American Physical Society},
	title = {Magnetic and all-optical switching properties of amorphous $\mathrm{Tb}_{x}\mathrm{Co}_{100-x}$ alloys},
	volume = {4},
	year = {2020}}

@article{HaidarNC2019,
	author = {Haidar, Mohammad and Awad, Ahmad A and Dvornik, Mykola and Khymyn, Roman and Houshang, Afshin and {\AA}kerman, Johan},
	journal = {Nature communications},
	number = {1},
	pages = {2362},
	publisher = {Nature Publishing Group UK London},
	title = {A single layer spin-orbit torque nano-oscillator},
	volume = {10},
	year = {2019}}

@article{Je2018,
	author = {Je, Soong-Geun and Rojas-S{\'a}nchez, Juan-Carlos and Pham, Thai Ha and Vallobra, Pierre and Malinowski, Gregory and Lacour, Daniel and Fache, Thibaud and Cyrille, Marie-Claire and Kim, Dae-Yun and Choe, Sug-Bong and Belmeguenai, Mohamed and Hehn, Michel and Mangin, St{\~A}{\copyright}phane and Gaudin, Gilles and Boulle, Olivier},
	doi = {10.1063/1.5017738},
	issn = {0003-6951},
	journal = {Applied Physics Letters},
	month = {02},
	number = {6},
	pages = {062401},
	title = {Spin-orbit torque-induced switching in ferrimagnetic alloys: Experiments and modeling},
	url = {https://doi.org/10.1063/1.5017738},
	volume = {112},
	year = {2018}}

@article{Mondal2023,
	author = {Sucheta Mondal and Debanjan Polley and Akshay Pattabi and Jyotirmoy Chatterjee and David Salomoni and Luis Aviles-Felix and Aur{\'e}lien Olivier and Miguel Rubio-Roy and Bernard Di{\'e}ny and Liliana Daniela Buda Prejbeanu and Ricardo Sousa and Ioan Lucian Prejbeanu and Jeffrey Bokor},
	issn = {0304-8853},
	journal = {Journal of Magnetism and Magnetic Materials},
	pages = {170960},
	title = {Single-shot switching in $\mathrm{Tb}$/$\mathrm{Co}$-multilayer based nanoscale magnetic tunnel junctions},
	volume = {581},
	year = {2023}}

@article{Iihama2018,
	author = {Iihama, Satoshi and Xu, Yong and Deb, Marwan and Malinowski, Gr{\'e}gory and Hehn, Michel and Gorchon, Jon and Fullerton, Eric E. and Mangin, St{\'e}phane},
	journal = {Advanced Materials},
	number = {51},
	pages = {1804004},
	title = {Single-Shot Multi-Level All-Optical Magnetization Switching Mediated by Spin Transport},
	volume = {30},
	year = {2018}}

@article{LAvilesSR2020,
	author = {Avil{\'e}s-F{\'e}lix, L. and Olivier, A. and Li, G. and Davies, C. S. and {\'A}lvaro-G{\'o}mez, L. and Rubio-Roy, M. and Auffret, S. and Kirilyuk, A. and Kimel, A. V. and Rasing, Th. and Buda-Prejbeanu, L. D. and Sousa, R. C. and Dieny, B. and Prejbeanu, I. L.},
	date = {2020/03/23},
	doi = {10.1038/s41598-020-62104-w},
	id = {Avil{\'e}s-F{\'e}lix2020},
	isbn = {2045-2322},
	journal = {Scientific Reports},
	number = {1},
	pages = {5211},
	title = {Single-shot all-optical switching of magnetization in Tb/Co multilayer-based electrodes},
	url = {https://doi.org/10.1038/s41598-020-62104-w},
	volume = {10},
	year = {2020}}

@article{Hansen1989,
	author = {Hansen, P and Clausen, C and Much, G and Rosenkranz, M and Witter, K},
	journal = {Journal of Applied Physics},
	number = {2},
	pages = {756--767},
	publisher = {AIP Publishing},
	title = {Magnetic and magneto-optical properties of rare-earth transition-metal alloys containing $\mathrm{Gd}$, $\mathrm{Tb}$, $\mathrm{Fe}$, $\mathrm{Co}$},
	volume = {66},
	year = {1989}}

@article{DenBroeder1991,
	author = {F.J.A. {den Broeder} and W. Hoving and P.J.H. Bloemen},
	doi = {https://doi.org/10.1016/0304-8853(91)90404-X},
	issn = {0304-8853},
	journal = {Journal of Magnetism and Magnetic Materials},
	pages = {562-570},
	title = {Magnetic anisotropy of multilayers},
	url = {https://www.sciencedirect.com/science/article/pii/030488539190404X},
	volume = {93},
	year = {1991}}

@article{Tsunashima2001,
	author = {S Tsunashima},
	doi = {10.1088/0022-3727/34/17/201},
	journal = {Journal of Physics D: Applied Physics},
	month = {aug},
	number = {17},
	pages = {R87},
	title = {Magneto-optical recording},
	url = {https://dx.doi.org/10.1088/0022-3727/34/17/201},
	volume = {34},
	year = {2001}}

@article{kim2022,
	author = {Kim, Se Kwon and Beach, Geoffrey SD and Lee, Kyung-Jin and Ono, Teruo and Rasing, Theo and Yang, Hyunsoo},
	journal = {Nature materials},
	number = {1},
	pages = {24--34},
	publisher = {Nature Publishing Group UK London},
	title = {Ferrimagnetic spintronics},
	volume = {21},
	year = {2022}}



\end{document}